\documentclass[usenatbib]{mnras}
\usepackage{natbib,amssymb,amsmath,times,graphicx,url,aas_macros}
\usepackage{setspace}      % To manually set line spacing
\usepackage{epstopdf}       % to convert eps images to pdf images
\usepackage{verbatim}       % to allow multiline comments
\usepackage{bm}		  % allow shorthand to bold text
\usepackage{fleqn}		  % flush left equations
\usepackage{Wurster_macros}	 % flush left equations

\newcommand\tnow{1500~yr}
  
\newcommand\model[4]{\text{#4}${M}_{#1}\beta_{#2}B_\text{#3}$}
\newcommand\imodel[3]{i${M}_{#1}\beta_{#2}B_\text{#3}$}
\newcommand\nmodel[3]{n${M}_{#1}\beta_{#2}B_\text{#3}$}
\newcommand\hmodel[2]{h${M}_{#1}\beta_{#2}$}
\newcommand\Bnx{$B_\text{-x}$}
\newcommand\Bpz{$B_\text{+z}$}
\newcommand\Bnz{$B_\text{-z}$}
\newcommand\Bpmz{$B_{\pm z}$}

\newcommand\betar[1]{$\beta_\text{r,0} = #1$}
\newcommand\betarz{$\beta_\text{r,0}$}
\newcommand\betars{$\beta_\text{r,0} = 0.005$}
\newcommand\betarf{$\beta_\text{r,0} = 0.02$}
\newcommand\Machz{$\mathcal{M}_\text{0}$}
\newcommand\Mach[1]{$\mathcal{M}_0 = #1$}

\defcitealias{WursterLewis2020sc}{Paper II}

\title[Non-ideal MHD vs Turbulence I: Discs]{Non-ideal magnetohydrodynamics vs turbulence I: \\ Which is the dominant process in protostellar disc formation?}

\author[Wurster \& Lewis]{James Wurster$^{1,2}$\thanks{jhw5@st-andrews.ac.uk} and Benjamin T. Lewis$^{3}$  \\
$^{1}$Scottish Universities Physics Alliance (SUPA), School of Physics and Astronomy, University of St. Andrews, North Haugh, St Andrews, Fife KY16 9SS, UK \\
$^{2}$School of Physics and Astronomy, University of Exeter, Stocker Rd, Exeter EX4 4QL, UK \\
$^{3}$School of Physics and Astronomy, Rochester Institute of Technology, Rochester N.Y. 14623, U.S.A.
}
\date{Submitted: Revised: Accepted: }
\pagerange{\pageref{firstpage}--\pageref{lastpage}} \pubyear{2020}

\begin{document}
\label{firstpage}
\bibliographystyle{mnras}
\maketitle

%Note: 250word count limit
%177 words
\begin{abstract}
Non-ideal magnetohydrodynamics (MHD) is the dominant process.  We investigate the effect of magnetic fields (ideal and non-ideal) and turbulence (sub- and transsonic) on the formation of circumstellar discs that form nearly simultaneously with the formation of the protostar.  This is done by modelling the gravitational collapse of a 1~M$_\odot$ gas cloud that is threaded with a magnetic field and imposed with both rotational and turbulent velocities.  We investigate magnetic fields that are parallel/anti-parallel and perpendicular to the rotation axis, two rotation rates and four Mach numbers.  Disc formation occurs preferentially in the models that include non-ideal MHD where the magnetic field is anti-parallel or perpendicular to the rotation axis.  This is independent of the initial rotation rate and level of turbulence, suggesting that subsonic turbulence plays a minimal role in influencing the formation of discs.   Aside from first core outflows which are influenced by the initial level of turbulence, non-ideal MHD processes are more important than turbulent processes during the formation of discs around low-mass stars.
\end{abstract}


%----------------------------------------------------------------------------------------------------------------
\section{Introduction}
\label{intro}

Molecular clouds that host star formation regions are host to many physical processes, including gravity, turbulence \citepeg{Larson1981,HeyerBrunt2004} and magnetic fields \citepeg{HeilesCrutcher2005,Crutcher2012}.  All of these processes are expected to play a role in the star formation process \citepeg{MaclowKlessen2004,MckeeOstriker2007,HennebelleChabrier2011,FederrathKlessen2012}.

Turbulent motions likely occur on all scales within a molecular cloud, where turbulence is supersonic on the cloud-scale \citepeg{Larson1981} and decays with decreasing spatial scales.  This typically yields cores that have subsonic motion \citepeg{Myers1983,JijinaMyersAdams1999,BerginTafalla2007}, although not in all cases; for example, some dense cores within the Serpens molecular cloud have supersonic motions \citep{WilliamsMyers1999}.  Independent of the level of turbulence, the gas motion in the cores may not be completely random, as many cores are observed to have motions that are consistent with uniform rotation \citepeg{Goodman+1993,Caselli+2002}, although more recent surveys are less frequently finding evidence for this organised rotation \citepeg{Tobin+2011,Tobin+2012}.   \citet{Goodman+1993} observed rotation rates in cores of $0.002 < \beta_\text{r} < 1.4$ and typical values of $\beta_\text{r} \sim 0.02$, where $\beta_\text{r}$ is the ratio of rotational energy to gravitational energy, while \citet{Caselli+2002} observed the lower range of $10^{-4}  < \beta_\text{r} <  0.07$ in their sample. 

The traditional view is that the gas motion in the dense, star forming cores is a superposition of random (i.e. turbulent) and coherent (i.e. rotational) motions.  Moreover,  turbulence and rotation are inextricably linked since turbulence can generate rotation. Numerical studies have shown that rotating cores can be formed from a molecular cloud that contains density perturbations but no initial velocity \citep{Verliat+2020}, or from a turbulent molecular cloud with initially smooth density and no coherent velocity profile \citepeg{GoodwinWhitworthWardthompson2004a,GoodwinWhitworthWardthompson2004b,Bate2012,Bate2018,WursterBatePrice2019}.

Turbulent motions necessarily become more complicated in the presence of magnetic fields.  For example, turbulent eddies become elongated in the direction of the magnetic field \citep{GoldreichSridhar1995}, and the power spectrum is typically shallower than  Kolmogorov \citepeg{Iroshnikov1963}.  Turbulent motion will also quickly tangle the magnetic field in ideal magnetohydrodynamics (MHD), since gas can only flow along the field lines; in this case, the magnetic field may become tangled enough to prevent further gas motion \citep[see the review by][]{HennebelleInutsuka2019}.  In order for turbulence to not prohibit gas motions, magnetic reconnection must occur to change the topology of the magnetic field and hence the assumption of ideal MHD must be relaxed.

Given its ubiquity, turbulence is often included when modelling the formation of stars \citepeg{MatsumotoHanawa2011,Seifried+2012,Seifried+2013,Joos+2013,Myers+2013,TsukamotoMachida2013,MatsumotoMachidaInutsuka2017,LewisBate2018,GrayMckeeKlein2018}.  The initial conditions of these models vary from low- ($M < 10$~\Msun) to high- ($M > 100$~\Msun) mass cores, sub- to super-sonic turbulence, and may be purely hydrodynamic or include magnetic fields.  While stars form in all these studies, the environments around them vary considerably, with some studies suggesting that the inclusion of turbulence promotes disc formation while other studies suggest that turbulence hinders it \citep[for a review, see][]{WursterLi2018}.  The former studies suggest that, since turbulence promotes disc formation, turbulence is a possible solution to the magnetic braking catastrophe \citep[i.e. in numerical models that include strong magnetic fields, circumstellar discs fail to form around young stars; e.g.][]{AllenLiShu2003,Galli+2006}. 

Molecular clouds are only weakly ionised \citepeg{MestelSpitzer1956,NakanoUmebayashi1986,UmebayashiNakano1990}, therefore ideal MHD is an incomplete description, even if simulations show that turbulence can prevent the magnetic braking catastrophe in the presence of ideal magnetic fields.  A more complete description requires non-ideal MHD \citepeg{WardleNg1999,Wardle2007}, where the important terms for star formation are Ohmic resistivity, ambipolar diffusion and the Hall effect. Previous star formation studies that included these three terms have shown that under certain magnetic geometries, circumstellar discs of 10s of au can form \citepeg{Tsukamoto+2015hall,WursterPriceBate2016,Tsukamoto+2017,WursterBatePrice2018hd,WursterBate2019}; like turbulence, non-ideal MHD has been proposed as a solution to the magnetic braking catastrophe.  It should be noted that these non-ideal MHD studies were performed in initially laminar gas flows.

Given that both non-ideal MHD and turbulence may promote disc formation under certain conditions, is one effect dominant and under what conditions is this true?  The 50~\Msun{} star cluster formation study of \citet{WursterBatePrice2019}, where the cloud was initialised with supersonic turbulence, concluded that discs formed around stars independent of initial magnetic field strength or inclusion/exclusion of non-ideal MHD.  This agreed with previous supersonic turbulence studies that turbulence promoted disc formation, even in strong magnetic fields \citepeg{Seifried+2013}.  This study further concluded that only turbulence -- and not non-ideal MHD -- was required for disc formation.  

Throughout the literature, turbulence has been observed and modelled on all scales, with varying conclusions.  The inclusion of non-ideal MHD (at least the combination of Ohmic resistivity, ambipolar diffusion and the Hall effect) is a more recent addition, and these processes need further investigation in conjunction with turbulent gas motions.  In this study, we investigate the competing effects of ideal/non-ideal MHD and sub/transssonic turbulence on the formation of discs around isolated, low-mass stars using a 3D self-gravitating, smoothed particle, radiative, non-ideal magnetohydrodynamics code.  In a companion paper, \citet{WursterLewis2020sc} (hereafter \citetalias{WursterLewis2020sc}), we investigate the effect these parameters have on the formation of the stellar core.  In Sections \ref{sec:methods} and \ref{sec:ic} of this paper, we summarise our methods and initial conditions, respectively.  In \secref{sec:results:discs}, we present our results and we conclude in \secref{sec:conclusion}.

%----------------------------------------------------------------------------------------------------------------
\section{Methods}
\label{sec:methods}

We solve the self-gravitating, radiation non-ideal magnetohydrodynamics equations using the three-dimensional smoothed particle hydrodynamics (SPH) code \textsc{sphNG}.  This code originated from \citet{Benz1990}, but has since been modified to include variable smoothing lengths \citep{PriceMonaghan2007}, individual time-stepping and sink particles \citep{BateBonnellPrice1995}, flux-limited diffusion radiative transfer \citep{WhitehouseBateMonaghan2005,WhitehouseBate2006}, magnetic fields \citep[for a review, see][]{Price2012}, and non-ideal MHD \citep{WursterPriceAyliffe2014,WursterPriceBate2016}.  For stability of the magnetic field, we use the source-term subtraction approach of \citet{BorveOmangTrulsen2001}, the constrained hyperbolic/parabolic divergence cleaning method of \citet{TriccoPrice2012} and \citet{TriccoPriceBate2016}, and the artificial resistivity as described in \citet{Phantom2018}.  For a more detailed description of our methods, see \citet{WursterBatePrice2018sd}.

We calculate the three non-ideal MHD coefficients -- Ohmic resistivity, ambipolar diffusion, and the Hall effect -- using version 1.2.3 of the \textsc{Nicil} library \citep{Wurster2016}.  For a review of non-ideal MHD, see \citet{Wardle2007} and \citet{BraidingWardle2012acc}.   We use the default parameters of \textsc{Nicil}, which includes the canonical cosmic ray ionisation rate of $\zeta_\text{cr} = 10^{-17}$~s$^{-1}$ \citep{SpitzerTomasko1968}, which contributes towards the ionisation of low-mass ions ($m = 2.31$~m$_\text{p}$, where m$_\text{p}$ is the proton mass), high-mass ions ($m = 24.3$~m$_\text{p}$), and dust grains.  The dust grains are comprised of a single species of radius $a_\text{g} = 0.1$~$\mu$m, bulk density $\rho_\text{bulk} = 3$~\gpercc{} and a dust-to-gas fraction of 0.01 \citep{Pollack+1994}, but are evolved as three populations: negatively, positively, and neutrally charged.
			
When \rhoxeq{-8} is reached late in the first hydrostatic core phase, a sink particle \citep{BateBonnellPrice1995} of radius $r_\text{acc} = 1$~au is inserted to allow us to investigate the formation and early evolution of the disc.  The sink size is similar to the minimum value recommended by \citet{MachidaInutsukaMatsumoto2014}, although we insert the sinks at a higher density. This this should minimise the numerical effects the sink particle will have on disc formation and its early evolution.  As gas particles enter this radius ($r_\text{acc} = 1$~au), they are accreted onto the sink if they pass the criteria as given in \citet{BateBonnellPrice1995}, and are automatically accreted if they come within 0.1~au of the sink particle.

There are several factors that contribute to the choice of timestep, including a particle's density, acceleration and signal velocity and well as criteria based upon  \textsc{sphNG}'s Runge-Kutta integrator; for a discussion on timesteps, see section 2.3.2 of \citet{Phantom2018}.
In most cases, the timestep is controlled by a Courant-Friedrichs-Lewy-like condition
\begin{subequations}
\label{eq:dt}
\begin{equation}
\text{d}t_\text{CFL} = 0.3\frac{h}{v_\text{sig}},
\end{equation}
where $h$ is the smoothing length of a particle and $v_\text{sig}$ is its signal velocity.  Naturally, as the gas gets more dense, shorter timesteps are required to resolve the motion since $h \propto \rho^{-1/3}$.  
Non-ideal MHD adds an additional timestep constraint of
\begin{equation}
\text{d}t_\text{nimhd} = \frac{h^2}{2\pi \max\left(\eta_\text{OR}, \left|\eta_\text{HE}\right|,\eta_\text{AD}\right)},
\end{equation}
\end{subequations}
where $\eta$ is the non-ideal MHD coefficient. In cases of high density and strong magnetic fields, $\text{d}t_\text{nimhd} \ll \text{d}t_\text{CFL}$ \citepeg{Maclow+1995,ChoiKimWiita2009,WursterPriceAyliffe2014}, which greatly affects how long we can evolve a simulation.
\textsc{sphNG} does not include super-time-stepping \citep{AlexiadesAmiezGremaud1996}, but does use implicit time-stepping for Ohmic resistivity as introduced in \citet{WursterBatePrice2018sd}.

%----------------------------------------------------------------------------------------------------------------
\section{Initial conditions}
\label{sec:ic} 

Our initial conditions are similar to our previous studies of low-mass star formation in a magnetised medium \citepeg{LewisBatePrice2015,LewisBate2017,LewisBate2018,WursterBatePrice2018sd,WursterBatePrice2018hd,WursterBatePrice2018ff}.  We embed a cold, dense sphere of mass $M =$ 1~\Msun{}, radius $R=4\times10^{16}$~cm, uniform density $\rho_0 = 7.43\times10^{-18}$~\gpercc{}, and initial sound speed $c_\text{s} = 2.2\times10^4$~\cms{} into a warm medium of edge length $L = 4R$; the cold sphere and warm medium are in pressure equilibrium and have a density contrast of 30:1.  The sphere is given a solid body rotation about the $z$-axis (i.e. $\bm{\Omega}_0 = \Omega_0\hat{\bm{z}}$), and the entire domain is threaded with a magnetic field of strength $B_0=1.63\times10^{-4}$~G = 163~$\mu$G, which is equivalent to five times the critical mass-to-flux ratio (i.e. \mueq{5}). 

A turbulent velocity field is added to the sphere, such that the total velocity of any given particle is a superposition of the initial rotational and turbulent velocities.  We calculate the turbulent component similarly to \citet{OstrikerStoneGammie2001} and \citet{BateBonnellBromm2003}, in which we generate a divergence-free %\footnote{i.e. a sinusoidal velocity field}
random Gaussian velocity field with a power spectrum $P(k) \propto k^{-4}$, where $k$ is the wavenumber.  %, on a $128^3$ uniform grid.  
The velocity field is normalised to obtain the desired Mach number and then the turbulent component of the particles' velocity is interpolated from this.  Each particle will have a different turbulent velocity, with the root mean square velocity of the turbulence given by $\bar{v}_\text{turb} = \mathcal{M} c_\text{s}$, where $\mathcal{M}$ is the desired Mach number.  This superposition of rotational and turbulent velocities means that our initial cores have different amounts of initial kinetic energy, where the increase in initial kinetic energy corresponds to the initial amount of turbulent energy.  For further details of how the initial velocity field is set up, see \citet{LewisBate2018}.  

Given the competing physical processes, it is useful to compare the various energies to the gravitational potential energy.  The useful ratios are
\begin{flalign}
\beta_\text{r} = E_\text{rot}/E_\text{grav} = \frac{1}{3}\frac{R^3\Omega^2}{GM}, \\
\beta_\text{turb} = E_\text{turb}/E_\text{grav} = \frac{5}{6}\frac{R\mathcal{M}^2 c_\text{s}^2}{GM}, \label{betaturb}\\
\beta_\text{mag} = E_\text{mag}/E_\text{grav} = \frac{5}{18}\frac{R^4B^2}{GM^2}, \label{betamag}\\
\alpha= E_\text{therm}/E_\text{grav} = \frac{5}{2}\frac{R c_\text{s}^2}{GM}, \label{alpha}
\end{flalign}
for rotational-to-gravitational\footnote{In Table 1 of \citet{LewisBate2018}, the values of $\Omega_0$ for $\beta_\text{r} \ge 0.01$ are incorrect; their $\beta_\text{r}$ correctly reflects their initial conditions.}, turbulent-to-gravitational\footnote{This differs from \citet{LewisBate2018}, which included a typo in their equation.}, magnetic-to-gravitational\footnote{This differs from \citet{LundBonnell2018} whose magnetic energy is for the sphere and surrounding region, assuming $B(r>R) = B_0(R/r)^3$; see \citet{Hartmann1998} for derivation and explanation.} and thermal-to-gravitational potential energy, respectively.  In this study, we use constant $\beta_\text{mag,0} = 0.071$ and $\alpha_0 = 0.36$\footnote{This value differs from \citet{LewisBate2018}, although their eqn.~32 yields this value rather than the value they present.}.

We include $10^6$ equal mass SPH particles in the sphere and an additional $5\times10^5$ particles in the warm medium.  

%----------------------------------------
\subsection{Parameter space}
We investigate the same parameter space both here and in \citetalias{WursterLewis2020sc}:
\begin{enumerate}
\item \emph{Magnetic processes}:  We investigate pure hydrodynamics, ideal MHD and non-ideal MHD.  All the non-ideal MHD models include Ohmic resistivity, ambipolar diffusion and the Hall effect.
\item \emph{Magnetic field direction}: For ideal MHD, we investigate the two directions of $\bm{B}_0 = -B_0\hat{\bm{x}} \equiv B_\text{-x}$ and $-B_0\hat{\bm{z}} \equiv B_\text{-z}$.  For non-ideal MHD, we investigate $\bm{B}_0 = -B_0\hat{\bm{x}}$, $-B_0\hat{\bm{z}}$ and $+B_0\hat{\bm{z}} \equiv B_\text{+z}$ since the Hall effect is dependent on the sign of $\bm{\Omega} \cdot \bm{B}$ \citepeg{BraidingWardle2012acc}.
\item \emph{Turbulent Mach number}: We investigate sub- and transsonic values of $\mathcal{M}_0 = $ 0, 0.1, 0.3 and 1.0.  These values correspond to ratios of turbulent-to-gravitational energy of $\beta_\text{turb,0} = 0$, $0.0012$, $0.011$ and $0.12$, respectively. \citet{LewisBate2018} found that due to the low mass of the core, supersonic values caused a large part of the cloud to unbind, preventing a useful investigation.  
\item \emph{Rotation}: We investigate rotation rates of $\Omega_0 = 1.77\times 10^{-13}$ and $3.54\times 10^{-13}$~s$^{-1}$, corresponding to ratios of rotational-to-gravitational energy of $\beta_\text{r,0} = 0.005$ and $0.02$, respectively; we call these slow and fiducial rotators, respectively.  The former matches the value used in our previous studies and the latter matches the peak of the observed distribution of rotation rates \citep{Goodman+1993}.
\end{enumerate}

Our magnetised models use the naming convention of \model{\emph{b}}{\emph{c}}{\emph{d}}{\emph{a}}, where $a$ = i (n) for ideal (non-ideal) MHD, $b$ is the Mach number, $c$ is the initial ratio of rotational-to-gravitational energy, and $d$ represents the orientation of the initial magnetic field (i.e. $\pm z$ or $-x$); our hydrodynamic models use the convention \hmodel{\emph{b}}{\emph{c}}.  An asterisk, *, in place of a variable indicates every model with the remaining defined components.
%----------------------------------------------------------------------------------------------------------------
\section{Results}
\label{sec:results:discs}

The models are evolved for at least 1500~yr after the sink particle is inserted (i.e. after the protostar has formed), which is at a different time for each model.  Although  1500~yr is a relatively short period in the lifetime of the disc \citep{Dunham+2014}, it allows us to investigate whether or not discs can form simultaneously with the formation of the protostar or during their very early evolution.  For a discussion about longer-term evolution, see \secref{sec:evol}.

\begin{figure*}
\centering
\includegraphics[width=0.45\textwidth]{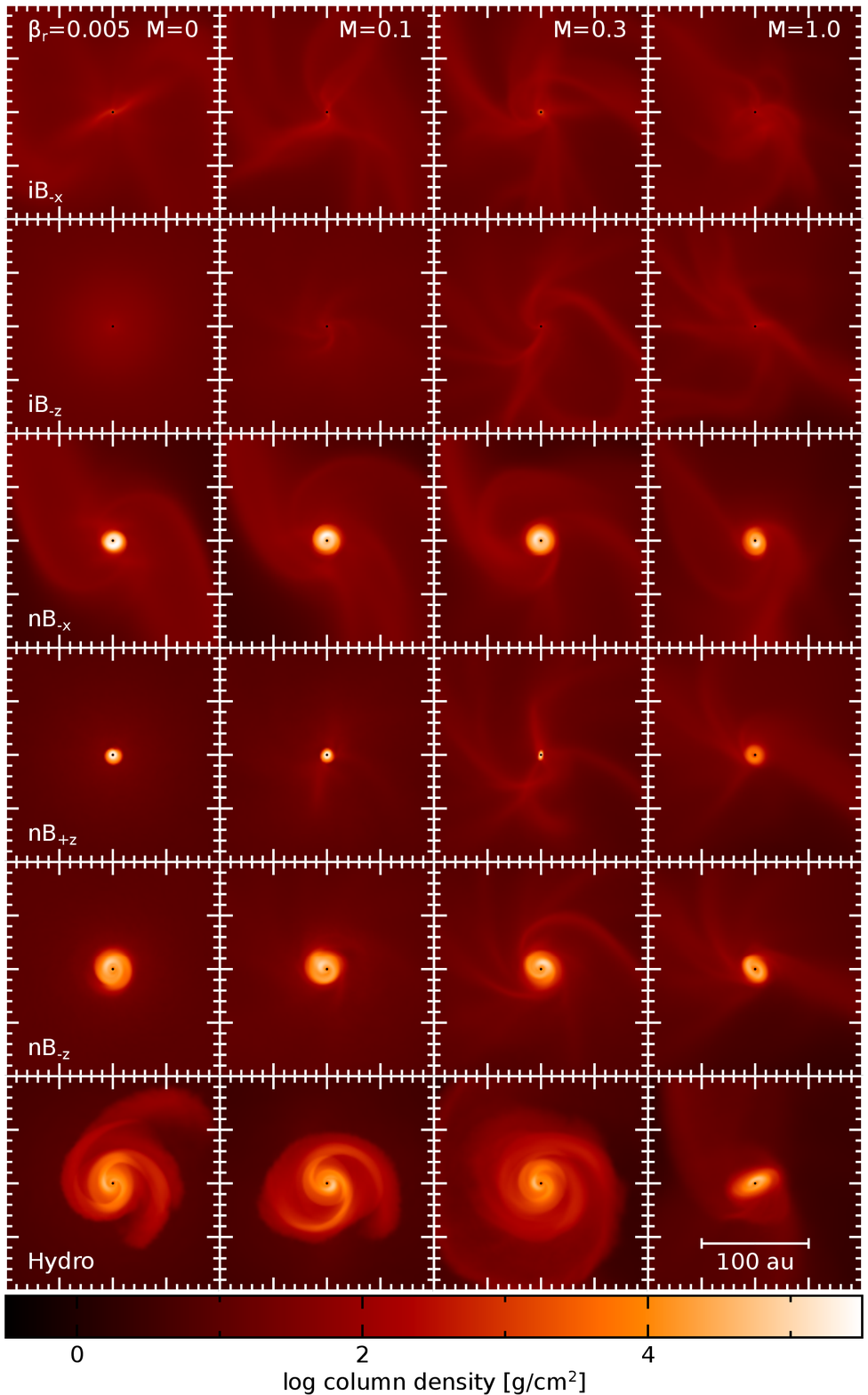}  %Made on DiAL
\includegraphics[width=0.45\textwidth]{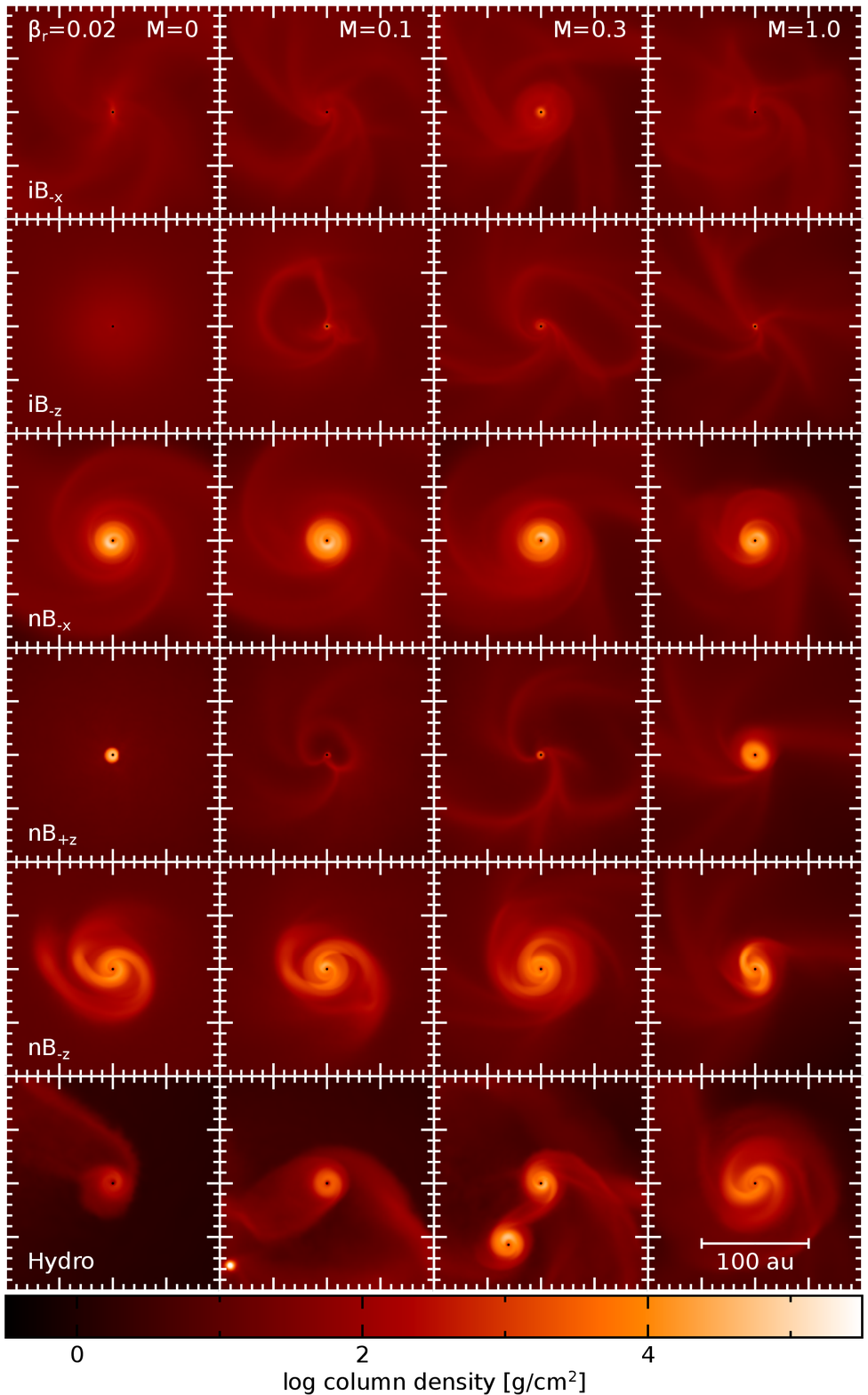}  %Made on DiAL
\caption{Vertically integrated gas column density for the models with \betar{0.005} (left) and 0.02 (right) at 1500~yr after the formation of the star.  The star has been shifted to be at the centre of each panel.  Models  \hmodel{0.0}{0.02}, \hmodel{0.1}{0.02} and \hmodel{0.3}{0.02} have fragmented into multiple stars, thus we have re-centred on the star with the largest disc.  The black dots represent the sink particles to scale.  In environments with sub/transsonic turbulence, magnetic fields are more important than turbulence in the formation of circumstellar discs. This is true for both slow rotators (left) and clumps that are initial rotating at the observed mean rate (right). }
\label{fig:rhogas}
\end{figure*}

\begin{figure*}
\centering
\includegraphics[width=0.45\textwidth]{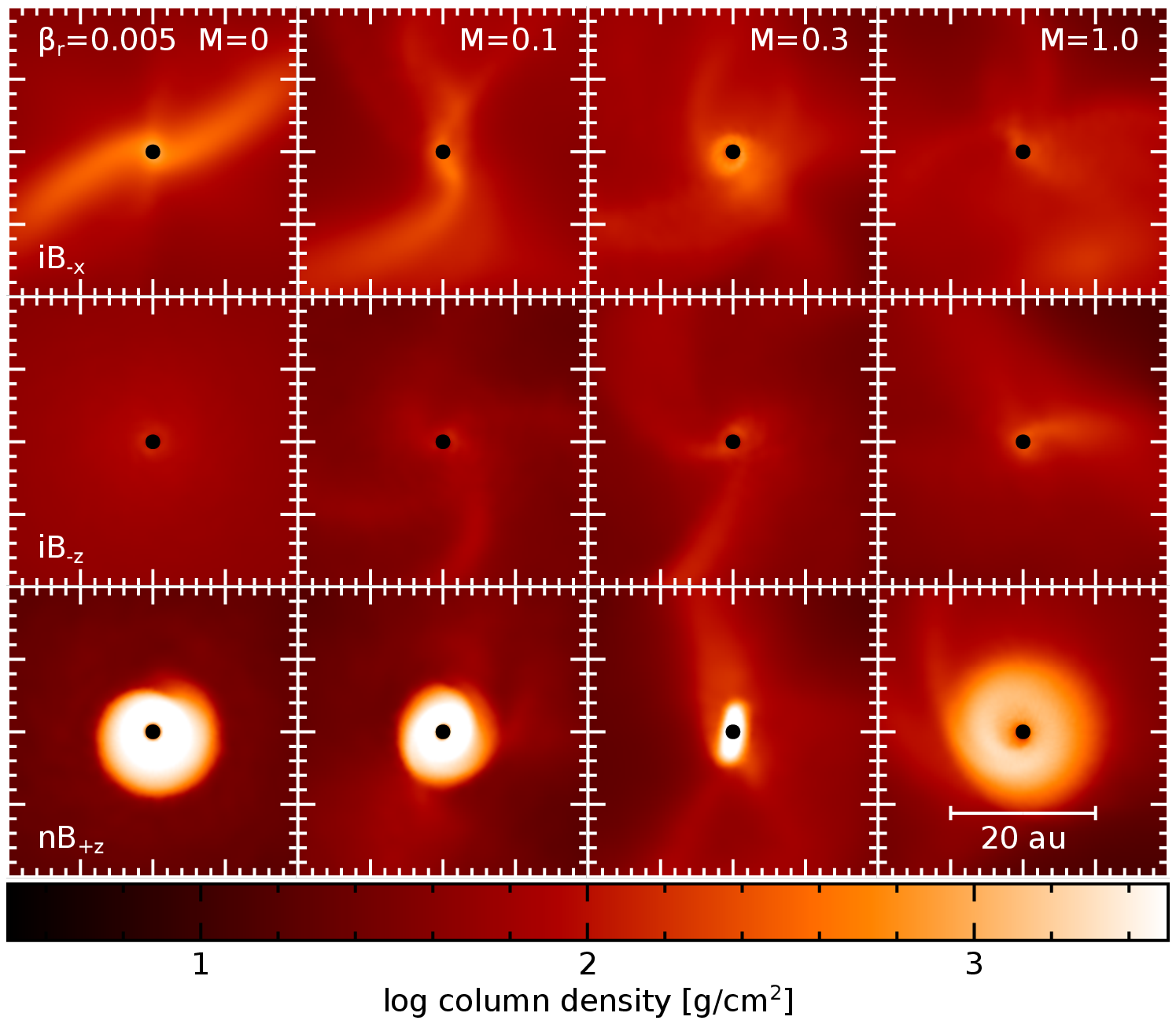}  %Made on DiAL
\includegraphics[width=0.45\textwidth]{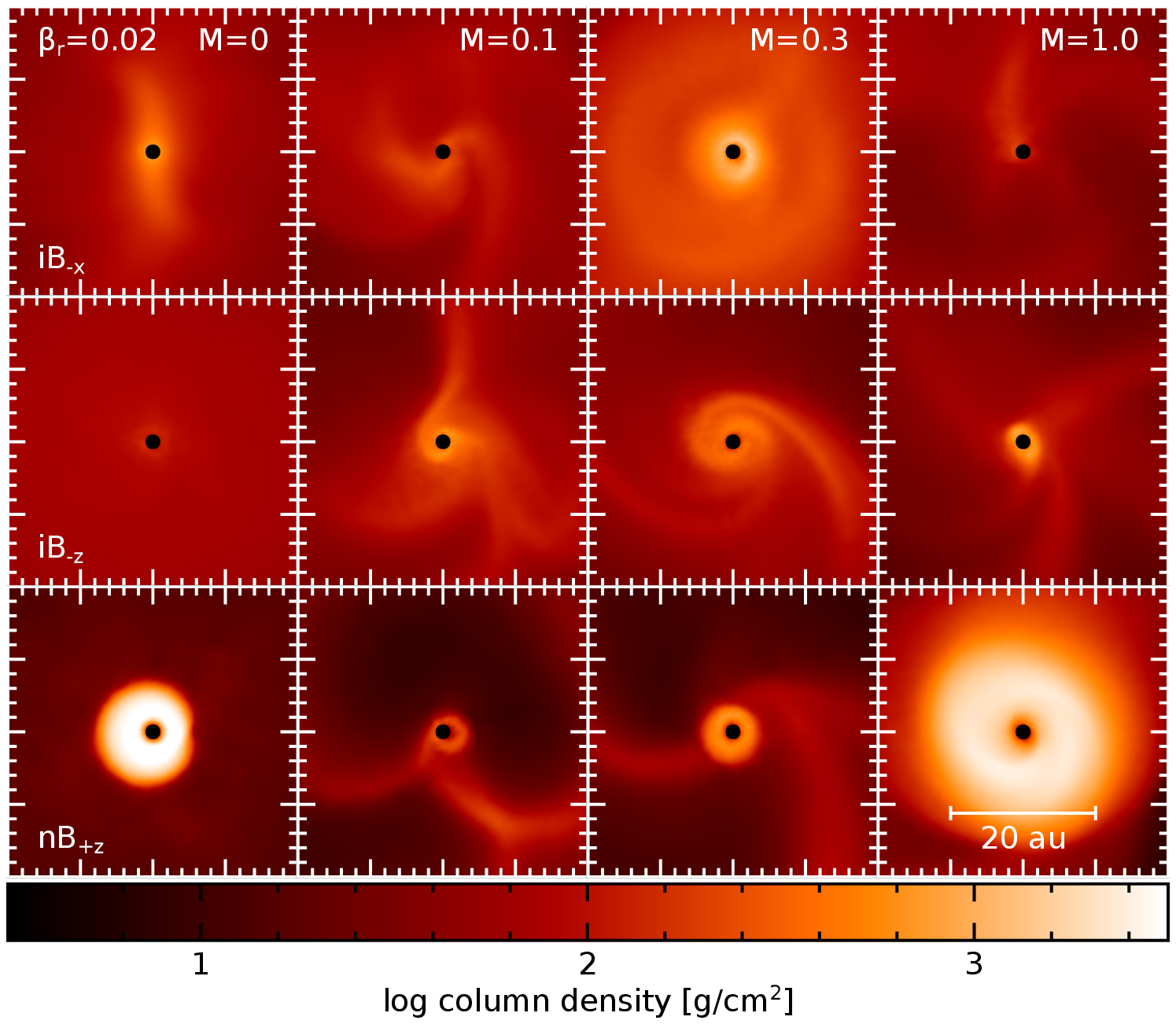}  %Made on DiAL
\caption{Vertically integrated gas column density as in \figref{fig:rhogas}, but zoomed in over a smaller region and only showing the models that form small or no discs.  Black dots represent the sink particles to scale.  Turbulence promotes the formation of small discs in  \imodel{>0}{0.02}{-z} and \imodel{0.3}{*}{-x}.}
\label{fig:rhogas:zoom}
\end{figure*}

\figref{fig:rhogas} shows the gas column density \tnow{} after the formation of the protostar for all the models in our suite, and \figref{fig:rhogas:zoom} shows the gas column density near the protostar for models that form small or no discs.  In each panel, the protostar has been shifted to the origin, but we have not altered the orientation of the discs.  

To quantitatively investigate the structure of a disc, we use the same method as presented in \citet{Bate2018} and \citet{WursterBatePrice2019}.  For each star, we find the closest particle and determine if it is bound to the star+disc system, has an eccentricity of $e < 0.3$ and has density \rhoge{-13}.  If so, it is added to the sink+disc system, and this process is repeated with the next nearest neighbour until all the particles have been checked.  The mass of the disc, $M_\text{disc}$, is the total mass of all particles that meet the criteria, and its radius $R_\text{disc}$ is the radius that contains 63.2 per cent of the disc mass; we only analyse discs with $M_\text{disc} > 10^{-4}$~\Msun.  This method of defining discs yields similar results to the method given in \citet{Joos+2013}.

\begin{figure}
\centering
\includegraphics[width=\columnwidth]{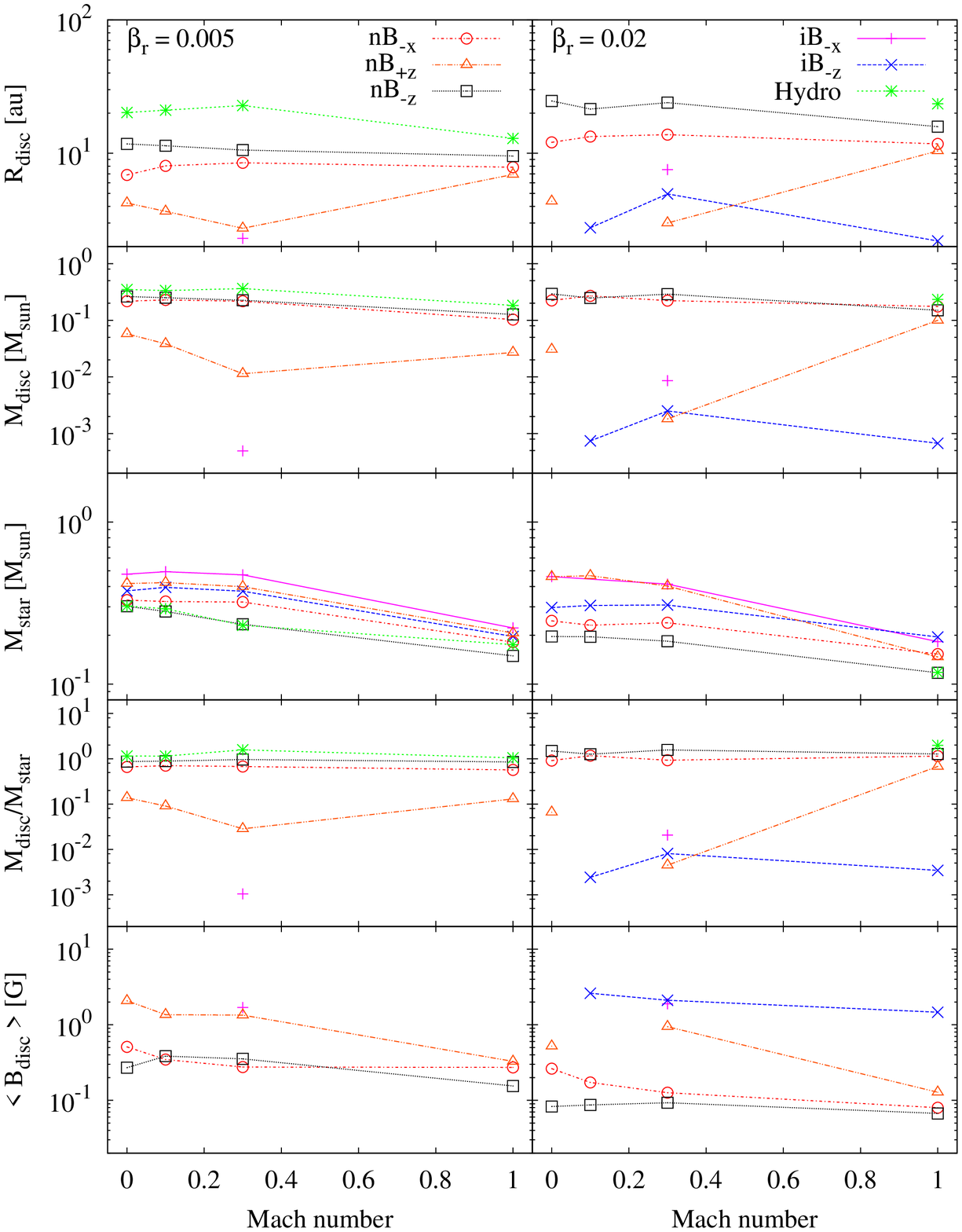}  %Made on DiAL
\caption{Disc radius, disc mass, stellar mass, disc-to-stellar mass ratio and the average magnetic field strength in the disc for each model \tnow{} after star formation; the hydrodynamic models with multiple stars are excluded. The legend is split across both columns for clarity.  For the models with well defined discs ($R_\text{disc} \gtrsim5$~au and $M_\text{disc} \gtrsim0.1$~\Msun{}), increasing the Mach number has minimal effect on the disc properties.}
\label{fig:rdisc}
\end{figure}

\figref{fig:rdisc} shows the radius, disc mass, stellar mass, disc-to-stellar mass ratio and average magnetic field strength for the disc in each model \tnow{} after the formation of the star.

%-----
\subsection{Discs}
\label{sec:discs}
\subsubsection{Hydrodynamics}
\begin{figure}
\centering
\includegraphics[width=\columnwidth]{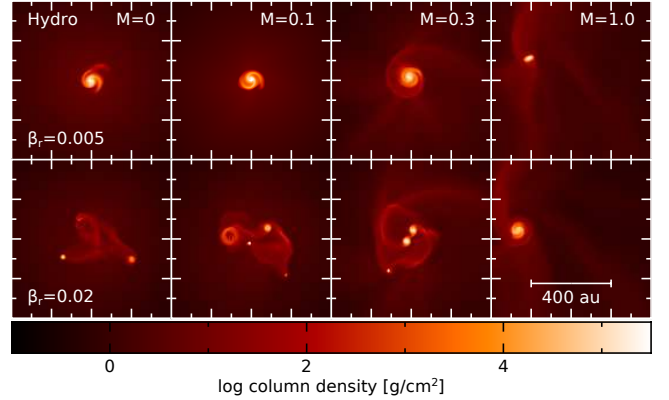}  %Made on DiAL
\caption{Vertically integrated gas column density for the hydro models.  The models have not been shifted nor rotated.  The transsonic models form single discs away from the origin, whereas the subsonic fiducial rotators fragment into multiple protostars, many of which have circumstellar discs. }
\label{fig:rhogas:hydro}
\end{figure}

Previous studies have shown that in the absence of magnetic fields and turbulence, large discs form \citepeg{Bate1998,SaigoTomisaka2006,SaigoTomisakaMatsumoto2008,MachidaInutsukaMatsumoto2010,Bate2010,Bate2011,BateTriccoPrice2014,WursterBatePrice2018hd}. This is true for our slow rotator models (\betars), which each form a single 10-30~au disc; these are slightly smaller than the \sm40~au discs predicted by \citet{Hennebelle+2016}, but typically agree within the factor of two range that  \citet{Hennebelle+2016} found.  There is no trend in disc size with Mach number, suggesting that rotation is more important than turbulence.   The three subsonic fiducial rotator models (\betarf) form higher order systems rather than a single, large disc.  %The transsonic model forms a 25~au disc.
\figref{fig:rhogas:hydro} shows the gas column density of the hydrodynamic models, zoomed out to encompass the entire system.  

In the higher order systems, the core fragments such that the protostars form nearly simultaneously.   In these models, there is initially more rotational than turbulent energy, suggesting that the fragmentation is a result of rotation alone.  This agrees with \citet{WursterBate2019}, who found that disc fragmentation could be induced from rotation alone if the initial rotational energy was large enough\footnote{In \citet{WursterBate2019}, the disc fragmented rather than the core itself since the initial core was centrally condensed as opposed to uniform, which promoted the formation of a single protostar, even under rapidly rotating initial conditions.}.  In the absence of rotation, \citet{GoodwinWhitworthWardthompson2004b} found that even low levels of turbulence ($\beta_\text{turb,0} > 0.01$) permitted fragmentation and multiple systems to form.  Therefore, in both extremes of rapid rotation or no rotation plus low levels of turbulence, fragmentation can occur to form multiple systems.  Between these two extremes, we find that slow rotation (\betar{0.005}) will stabilise the core against fragmentation and permit a single disc to form.  Therefore, rotational energy is more important than subsonic turbulent energy for determining the stability of hydrodynamic discs.

Both transsonic models form a single disc, and, unlike the subsonic models, these discs are not at the origin of system.  Thus, at least transsonic turbulence is required to disrupt the gas flow such that the disc forms away from the origin, independent of rotational energy.  In our fiducial rotation, transsonic model (\hmodel{1.0}{0.02}), $\beta_\text{turb,0}  > \beta_\text{r,0}$ suggests that transsonic turbulence can stabilise against the fragmentation of a rapidly rotating disc.  This may be a result of the turbulent motions creating gas flows away from the origin, which in turn reduces the importance of the rotational energy about the origin.

%-----
\subsubsection{Ideal MHD}
From our initial conditions, $\beta_\text{mag,0} > \beta_\text{r,0}$ for all magnetised models, and $\beta_\text{mag,0} > \beta_\text{turb,0}$ for all subsonic models.  Thus, magnetic fields are expected to play an important role in the evolution of the cloud and subsequent formation of the disc.

Including ideal magnetic fields hinders disc formation, preventing discs from forming in 11 of 16 models; see \figref{fig:rhogas:zoom}.  The only slow rotator to form a disc is \imodel{0.3}{0.005}{-x}, which forms a low-mass, $r \approx2.3$~au disc; for comparison, its fiducial rotator counterpart forms a 7.5~au disc.  Since these are the only ideal models with \Bnx{}  to form discs, their formation is possibly a result of the random seed number used to generate the turbulence \citepeg{GoodwinWhitworthWardthompson2004a,Liptai+2017,Geen+2018} and a fortuitous combination of the other initial conditions.  The turbulent fiducial rotators with \Bnz{} form discs with $r < 5$~au.

Ideal models with \Bpmz{} are most susceptible to the magnetic braking catastrophe \citepeg{AllenLiShu2003,Galli+2006}, thus the consistency of disc formation in \imodel{> 0}{0.02}{-z} suggests that turbulence combined with reasonable rotation can hinder the magnetic braking catastrophe to permit small discs to form.  However, these discs are much smaller than those inferred around Class 0 objects \citepeg{Dunham+2011,Tobin+2015}, thus turbulence appears to weaken rather than prevent the catastrophe.  In strong magnetic fields ($\mu_0 \lesssim 5$), \citet{WursterBate2019} found that the more massive discs were formed in laminar models with \Bnx{} rather than \Bnz{}.  Therefore, turbulence appears to change this result, promoting disc formation in the \Bnz{} models rather than the \Bnx{} models; again, we caution that this conclusion may be a result of the seed for our initial turbulent velocity field.

%-----
\subsubsection{Non-ideal MHD}
Ideal magnetic fields efficiently transport angular momentum from the centre of the collapsing cloud outwards, hindering disc formation (i.e. the magnetic braking catastrophe).  In \imodel{> 0}{0.02}{-z} discussed above, turbulence can hinder the angular momentum transport enough to permit a small, $\lesssim$~5~au disc to form.  Another solution to preventing angular momentum transport is non-ideal MHD, in which interactions between neutral gas and ions weaken and reshape the magnetic field.  When including non-ideal MHD in models of laminar gas flows, large discs are expected to form in magnetic fields of \Bnz{} and small discs for \Bpz{} \citep[][and was confirmed numerically by \citealp{Tsukamoto+2015hall,WursterPriceBate2016,WursterBatePrice2018hd}]{BraidingWardle2012acc}.

A single disc forms in every non-ideal MHD model except \nmodel{0.1}{0.02}{+z}, although this outlier may be a result of the random seed of the turbulence.  The non-ideal models with \Bnx{} and \Bnz{} have disc properties (radius, mass and disc-to-stellar mass ratio) that are insensitive to the level of turbulence, with each property varying by less than a factor of two as the initial Mach number is increased from zero to one.  Although turbulence may hinder the transport of angular momentum in these models, its effect is clearly secondary to that of including non-ideal processes.  In all cases, the discs in the fiducial rotators (\betarf) are larger and more massive than their slow rotator counterparts.  This suggests that sub- and transsonic turbulence plays a subordinate role (even to rotation) in disc formation when including non-ideal MHD at these orientations.

The Hall effect hinders disc formation for \Bpz{}, and this is clearly shown in \figsref{fig:rhogas}{fig:rhogas:zoom} where discs of $r \lesssim 10$~au form.  These models may be slightly more sensitive to the initial Mach number than the other initial configurations, with disc radius and mass initially decreasing before increasing again for \Mach{1}.  The largest discs in this configuration appear in the transsonic models, although these discs are only twice as large as their laminar counterparts.  This suggests that turbulence contributes to disc formation in this case, although this may be dependent on the initial structure of the turbulence field, as discussed above.

%----------------------------------------
\subsection{Stellar mass}
The stellar masses at the end of the simulations are given in the third row of \figref{fig:rdisc}.  The models where disc formation is hindered (i.e. ideal MHD and to a lesser extent \nmodel{*}{*}{+z}) have higher stellar masses since the gas has less angular momentum and is more easily accreted onto the protostar.  Increasing the initial rotation rate also decreases the final stellar mass since more angular momentum of the gas must be shed prior to it being accreted onto the protostar.  Finally, increasing the initial level of turbulence decreases the final stellar mass, since the turbulent motions also hinder the gas flow towards the star.

The final stellar mass is intrinsically linked to the presence and magnitude of the processes which hinder angular momentum transport (e.g. adding non-ideal MHD, increasing \Machz{} or \betarz{}).  As additional processes are added, or the effect of any process is increased, the mass of the resulting star decreases.

%----------------------------------------
\subsection{Magnetic fields}

\begin{figure*}
\centering
\includegraphics[width=0.45\textwidth]{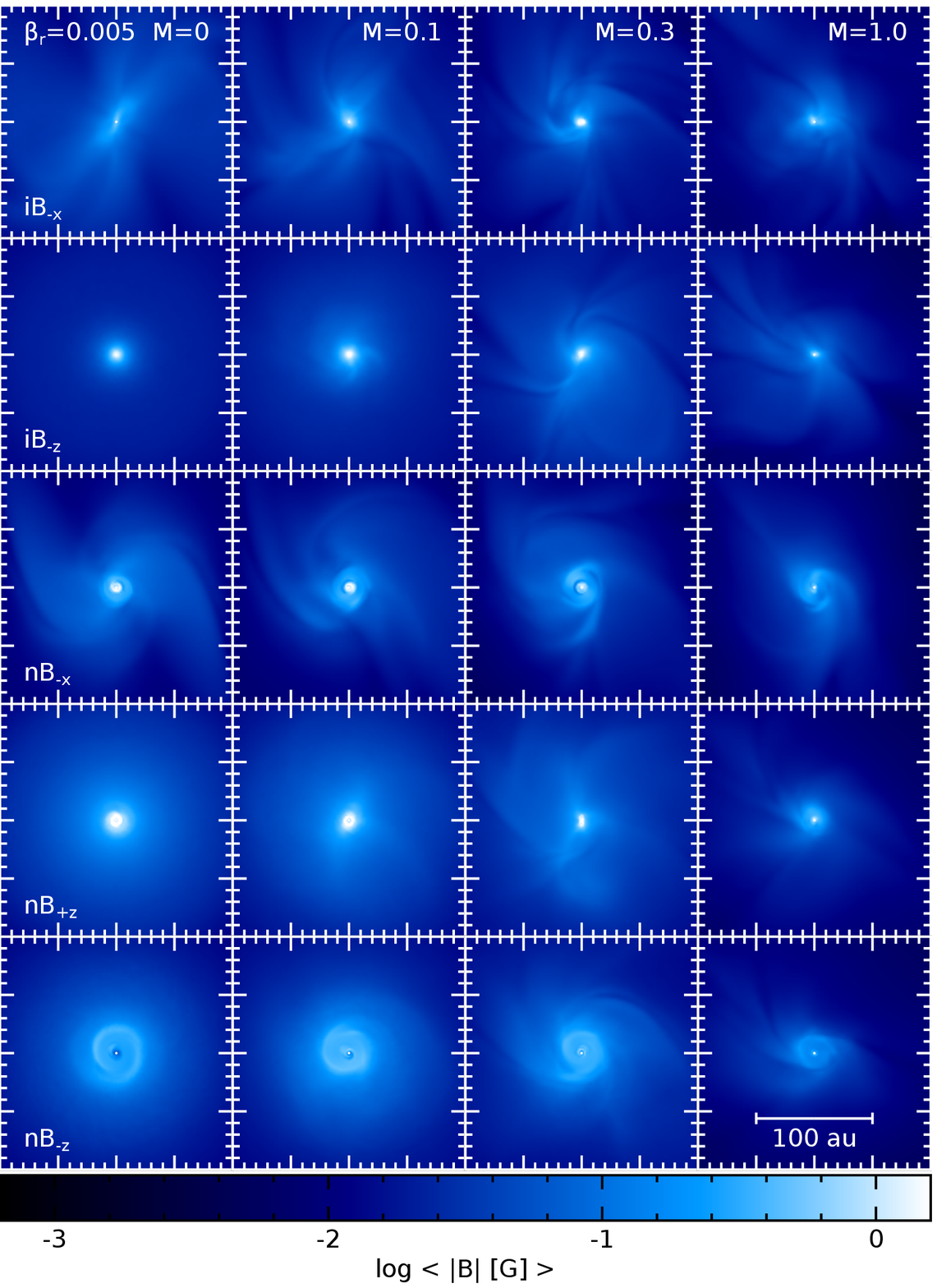}  %Made on DiAL
\includegraphics[width=0.45\textwidth]{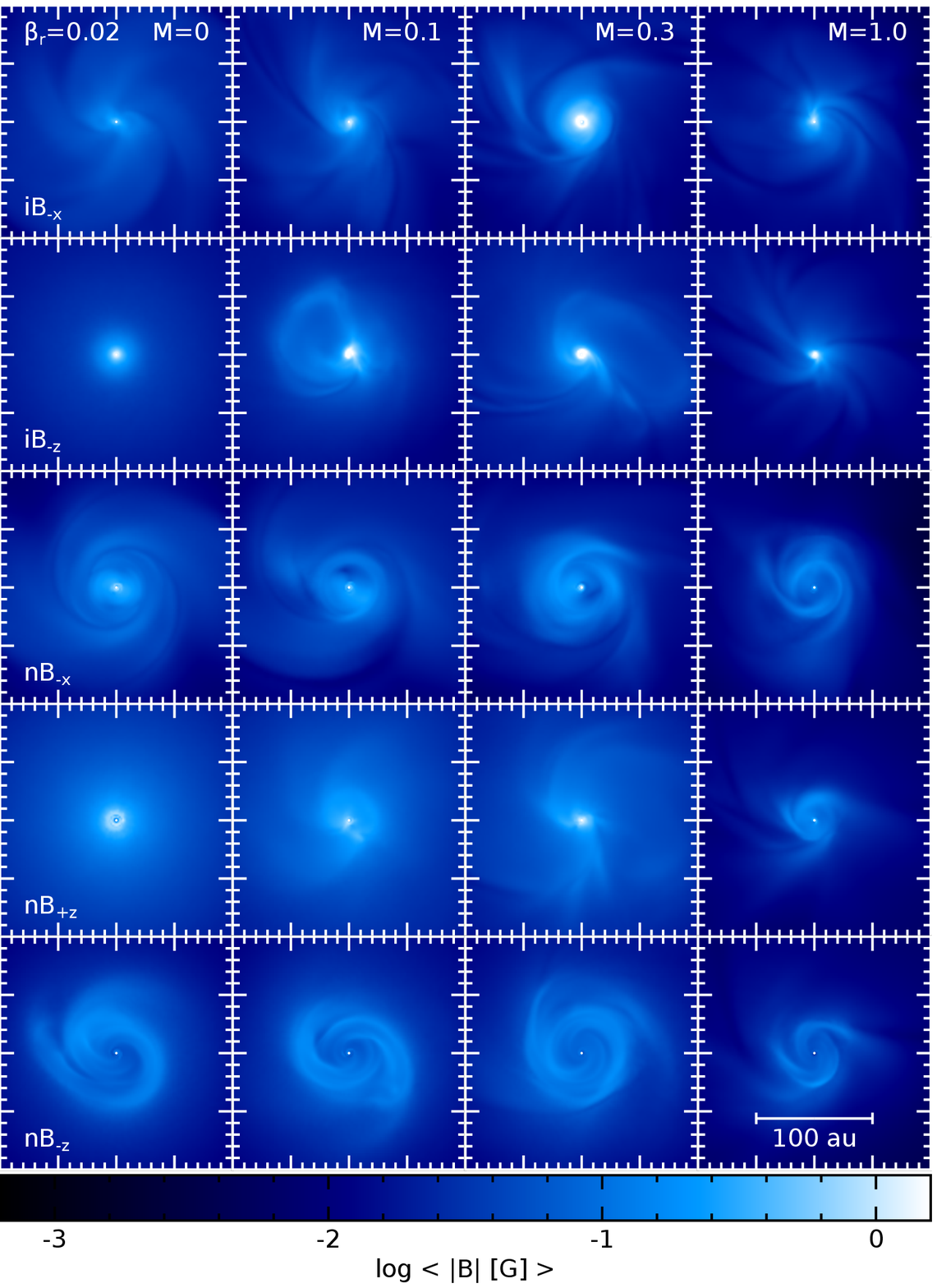}  %Made on DiAL
\caption{Density-weighted line-of-sight averaged magnetic field strength along the $z$-axis for the models as in \figref{fig:rhogas}.  Magnetic field strengths are generally unaffected by the turbulence, with the maximum magnetic field strength changing by a factor of \sm2 when all initial conditions except Mach number are held constant.  Magnetic field strengths are much weaker for the non-ideal models than the ideal models, and are generally weaker for faster initial rotations. }
\label{fig:B}
\end{figure*}

The magnetic field strength around the stars is shown in \figref{fig:B} for each of our magnetised models, and the average field strength in each disc is shown in the bottom panels of \figref{fig:rdisc}.
Changing the initial Mach number has minimal affect on the maximum field strength in all the magnetised models, with the maximum strength changing by only a factor of $\lesssim 3$ when all other initial properties are kept the same.  Greater changes occur when changing each of the other initial parameters, especially by including the non-ideal processes which disperse and diffuse the magnetic field.  

Increasing the initial rotation rate or including non-ideal MHD with \Bnz{} increases the angular momentum in the first hydrostatic core (FHC; see \secref{sec:fhc} below), even if only slightly.   This is enough to trap the magnetic field at slightly larger radii, which prevents the field lines from piling up and causing a strong magnetic field as seen in the other magnetised models.  Thus, the resulting magnetic fields are weaker in the fiducial rotators and in \nmodel{*}{*}{-z}.  Although the turbulence in these models affects the structure of the magnetic field, it is not strong enough to have a significant affect on the field strength itself.  

When considering the magnetic field strengths in the discs of the non-ideal models with \Bnx{} and \Bnz{} (bottom panels of \figref{fig:rdisc}), the average value changes by a factor of $\lesssim 3$ as the initial Mach number changes; there appears to be a slight trend of decreasing strength for higher Mach numbers, but this change is too small to be conclusive.  The non-ideal models with \Bpz{} show a larger variation in magnetic field strength, however, the small discs produced render it difficult to make a firm conclusion on any notable trends.  These results suggest that the initial magnetic properties have a dominant role over subsonic turbulence in setting the magnetic field strength in the disc. 

%----------------------------------------
\subsection{First hydrostatic core}
\label{sec:fhc}
\subsubsection{Angular momentum}
\label{sec:rotrate}
\begin{figure}
\centering
\includegraphics[width=\columnwidth]{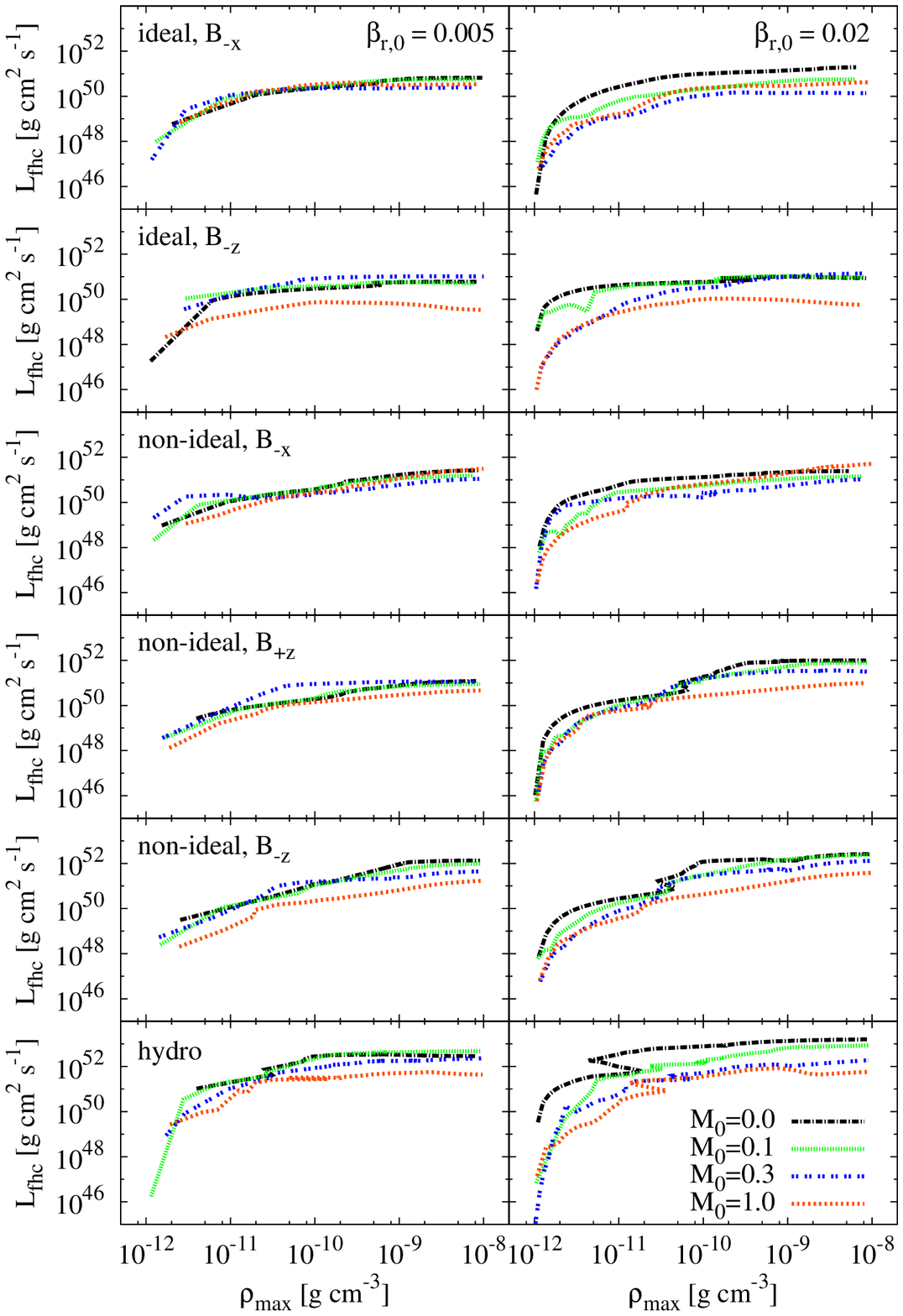}  %Made on DiAL
\caption{Angular momentum in the first hydrostatic core from its formation until the formation of the protostar (i.e. until the insertion of the sink particle) as function of maximum density, which is a proxy for time.  The first hydrostatic core is defined to be composed of all the gas with \rhoge{-12}.  Generally, increasing the initial Mach number decreases the angular momentum in the first hydrostatic core.  In agreement with previous studies, \hmodel{*}{*} and \nmodel{*}{*}{-z} have the largest angular momenta in the suite of simulations.}
\label{fig:Lfhc}
\end{figure}

Magnetic fields affect the angular momentum of the FHC as it forms and evolves \citepeg{Tsukamoto+2015oa,Tsukamoto+2015hall,WursterBatePrice2018hd} since they efficiently transport angular momentum outward into the lower density gas.  As a result, the ideal models have the lowest angular momentum in their FHCs while the hydro models have the highest.  This is shown in \figref{fig:Lfhc} where, as in our previous studies, we define the FHC to be comprised of all the gas with $\rho \ge 10^{-12}$~\gpercc.

Generally, in our models, increasing the initial level of turbulence decreases the angular momentum in the FHC; this decrease can be up to a factor of \sm100 in some cases.  Therefore, on the scale of our simulations, turbulence does not generate rotation.  This reduction in angular momentum for increasing turbulence does not necessarily affect the resulting disc properties as discussed in \secref{sec:discs}, which are insensitive to the initial levels of turbulence.  

Given the values presented in \figref{fig:Lfhc}, the angular momentum of the FHC alone cannot dictate whether or not it will fragment into multiple stars since a few models that form single stars have more angular momentum than the FHCs that fragment.  Nor can the angular momentum of the FHC alone dictate the properties of the resulting discs.  Therefore, other factors including magnetic fields and their orientation \citepeg{LewisBate2017} must be considered when determining if a core will fragment and what will be the properties of the resulting discs.

%-----
\subsubsection{Outflows}
\begin{figure*}
\centering
\includegraphics[width=0.45\textwidth]{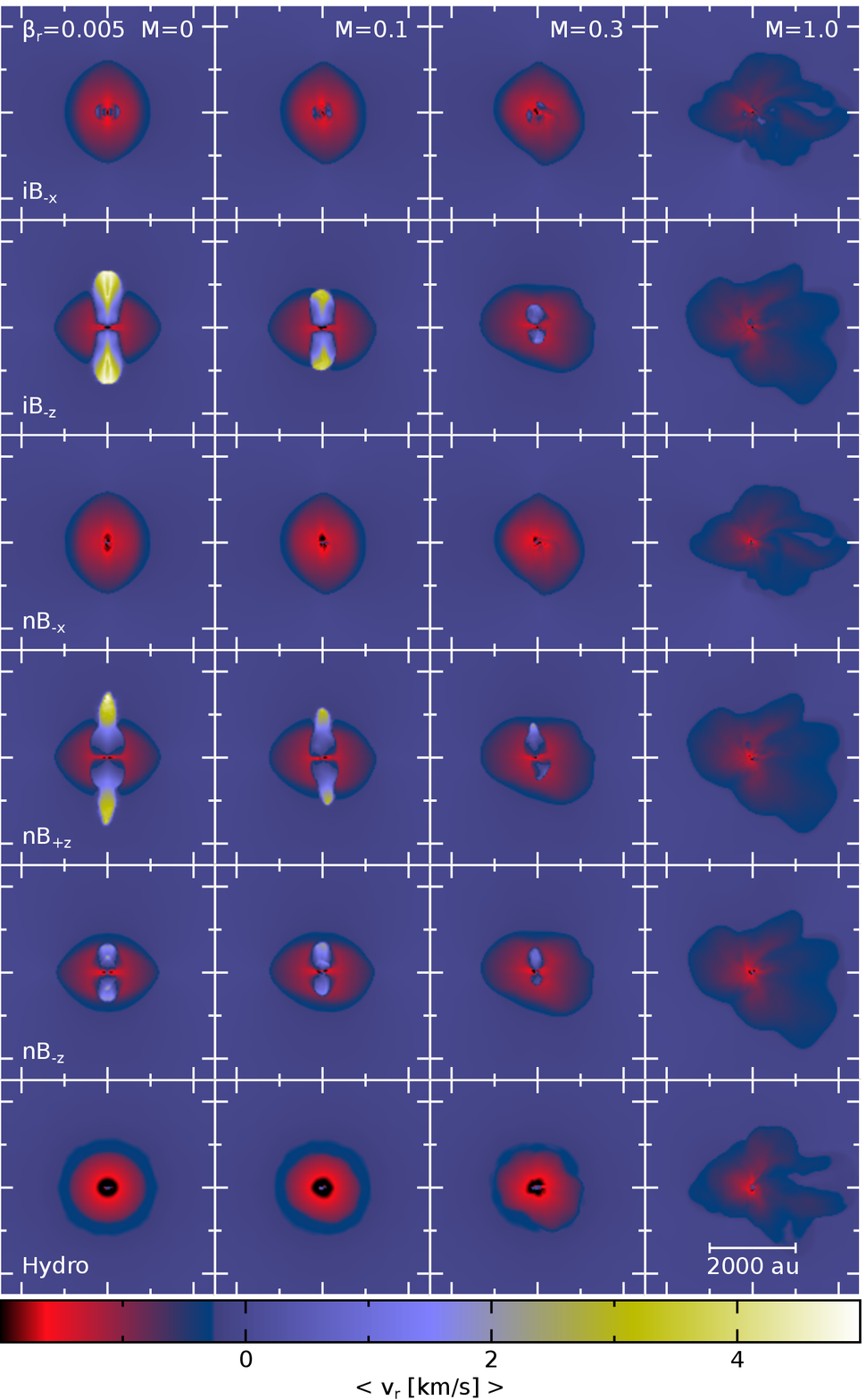}  %Made on DiAL
\includegraphics[width=0.45\textwidth]{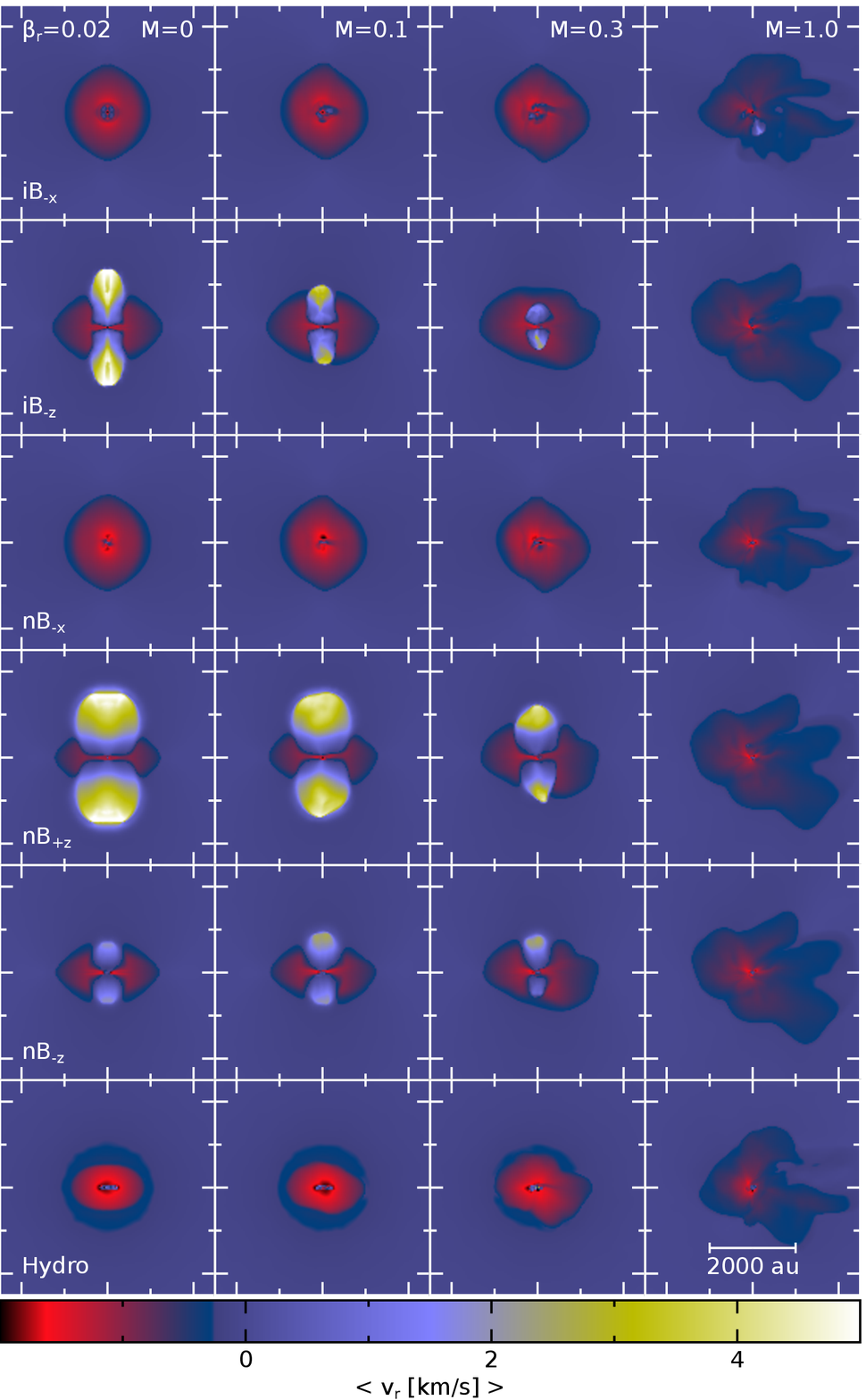}  %Made on DiAL
\caption{Density-weighted line-of-sight averaged radial velocity for the models as in \figref{fig:rhogas}.  The magnetically launched outflows are suppressed in the models with \Bnx{}.  Increasing the initial Mach number suppresses the outflows by tangling the magnetic field at the launching radius near the star and by creating a disordered gas motion in the path of the outflow. }
\label{fig:Vr}
\end{figure*}

 \figref{fig:Vr} shows the radial velocity for each model in our suite. 
 \citet{LewisBate2018} showed that increasing the initial Mach number in ideal models with \Bpz{} (the only orientation they studied) distorted and ultimately prevented first core outflows.  Our models with \Bpmz{} agree with this trend.  In agreement with our previous work, there are no outflows in the models with \Bnx{}.
As the initial Mach number increases, the outflows are slower and progress to a shorter distance in a similar amount of time since star formation and are typically less broad.  For the transsonic models, outflows are completely suppressed.

Unlike the previous characteristics we have investigated in this paper, the first core outflows are influenced by all initial conditions, including the initial level of turbulence.  These first core outflows are magnetically launched, so are not present in the hydro models; they are suppressed in the \Bnx{} models since a strong, poloidal magnetic field cannot be created near the star.  Fast outflows are launched from regions of lower ratios of toroidal-to-poloidal magnetic field components \citepeg{BateTriccoPrice2014,WursterBatePrice2018sd,WursterBatePrice2018hd}, which occur in the non-ideal models with \Bpz{} since the Hall effect decreases the toroidal component above and below the protostar.  The outflows are further suppressed as the initial turbulent amplitude is increased since the outflow encounters less well-ordered gas as it expands and since the magnetic field near the star (i.e. at the launching radius) becomes tangled and cannot generate a coherent outflow.  At the higher initial rotation rate, the outflows are more broad and slightly faster due to the additional initial angular momentum.  

%----------------------------------------
\subsection{Resolution}
Our initial spheres contain 1~\Msun{} of gas and $10^6$ particles, thus the Jeans mass is resolved throughout the collapse \citep{BateBurkert1997}.  At 1500~yr, there are $\gtrsim10^{4-5}$ particles in the discs that have formed, with particle smoothing lengths of $0.2-1$~au.  

A disc is considered resolved if it resolves the Toomre-Mass and the scale-height, $H$.  From this, \citet[][modified by \citealp{WursterBate2019} to account for MHD]{Nelson2006} suggested that the disc was resolved if
\begin{equation}
\label{eq:TM}
\Sigma < \Sigma_\text{reso} = \frac{\pi \left(c_\text{s}^2+v_\text{A}^2\right)^2}{G^2 m_\text{sph}N_\text{reso}}
\end{equation}
and
\begin{equation}
\frac{H}{h_\text{mid-plane}} = \frac{\sqrt{c_\text{s}^2+v_\text{A}^2}}{h_\text{mid-plane}\Omega} > 4,
\end{equation}
where $\Sigma$ is the surface density of the disc, $c_\text{s}$ is the sound speed, $v_\text{A}$ is the \alfven{} velocity, $G$ is Newton's gravitational constant, $m_\text{sph}$ is the mass of the SPH particle, $N_\text{reso}\approx  342$ is the number of particles required to resolve the maximum surface density for a cubic spline smoothing kernel, and $h_\text{mid-plane}$ is the smoothing length in the midplane of the disc.

In our models, the surface density of the discs that form is a few orders of magnitude lower than $\Sigma_\text{reso}$ thus they meet the resolution requirement of the Toomre-Mass.  In the discs, $H/\bar{h} \sim 6-8$, where $\bar{h}$  is the average smoothing length within $20^\circ$ of the mid-plane at any given radius.  Since $\bar{h} > h_\text{mid-plane}$, then $H/h_\text{mid-plane} > H/\bar{h}$, and we can conclude that our discs are resolved.

%----------------------------------------
\subsection{Evolutionary period}
\label{sec:evol}
Due the large suite presented here and in our companion paper \citepalias{WursterLewis2020sc}, the resolution of our models and our limited computational resources, we evolve the systems for $\gtrsim$1500~yr after the protostar formed, which is a very short phase in the lifetime of a disc, where the Class 0 phase alone is expected to last for $\sim 1.6\times10^5$~yr \citep{Dunham+2014}.  

Numerically, the lack of reasonable disc in the ideal models yields a reasonable minimum CLF timestep, d$t_\text{CFL}$ (recall Eqn.~\ref{eq:dt} and associated discussion), thus these models could be evolved for a reasonable length of time after disc formation.  In the non-ideal models, the CFL timestep is approximately 10 times smaller.  However, the presence of the dense disc and the non-ideal effects yield non-ideal timesteps d$t_\text{nimhd}$ that are a factor of \sm$10^3$ smaller than its CLF condition (i.e. \sm$10^4$ smaller than the CFL condition required in the ideal models); thus, running several non-ideal models as we do here severely limits the length of time we can evolve them.  So that we can perform a direct comparison amongst the models, we have intentionally analysed all models at the shortest end time of 1500~yr.
%Hydro CFL is approx 0.5dex lower than ideal CFL

\begin{figure*}
\centering
\includegraphics[width=\textwidth]{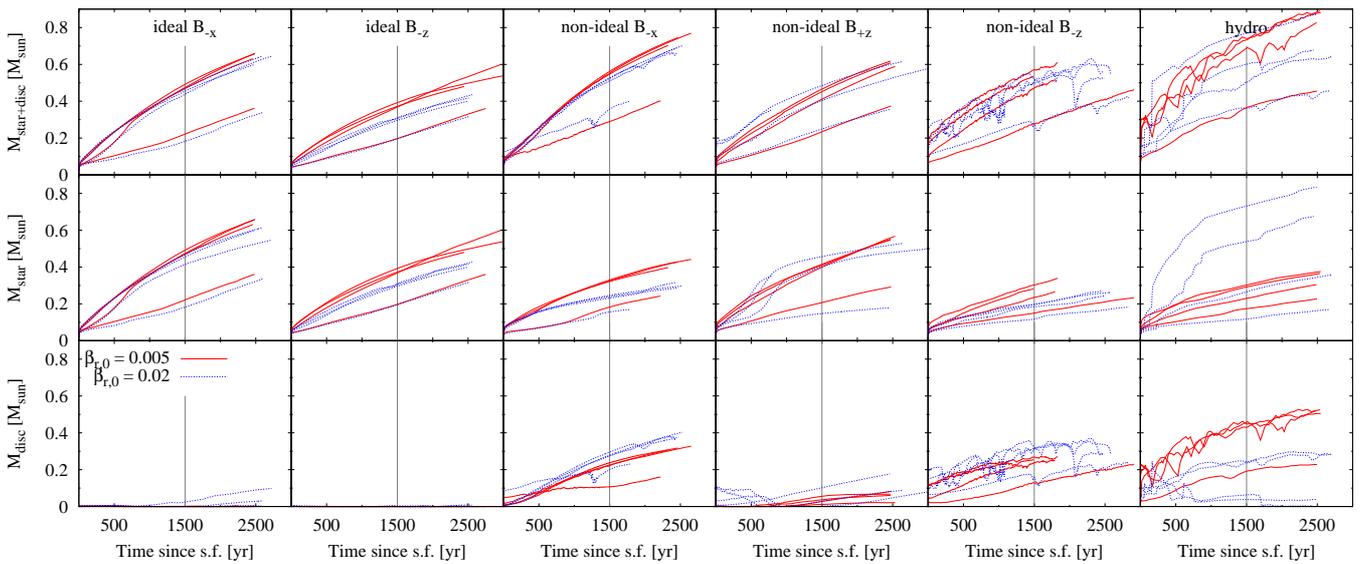}  %Made on DiAL
\caption{The time evolution of the amount of gas in the star+disc system (top row), the star itself (middle row) and in the disc (bottom row) for all our models.  We do not distinguish between the various Mach numbers since this is to show general trends only.  The vertical line at 1500~yr represents the time of analysis in this paper.  At 1500~yr, there is between 0.18 and 0.77~\Msun{} of gas in the star+disc system, and the masses continue to increase with time.}
\label{fig:Mevol}
\end{figure*}

The limited evolution allows us to reach conclusions about the disc formation that occurs simultaneously with protostar formation, which is likely the seed for any later evolution.  Since 0.18-0.77~\Msun{} of gas is in the star+disc system by 1500~yr, there is still a considerable amount of gas remaining in the initial core that could permit later evolution.  \figref{fig:Mevol} shows the mass of the star+disc system, stellar mass and disc mass as a function of time, normalised to the formation time of the protostar.  This shows that the envelope is continuing to collapse onto the star+disc system, and that the existing stars and discs are continuing to gain mass.  Although limited in time, this also shows that discs form simultaneously with the protostar formation rather than delayed by a few hundred or thousand years since $M_\text{disc}(t=0) > 0$; this is in agreement with \citet{WursterBatePrice2018hd}.

The discs and the envelope undergo long-term evolution, as shown by \citet{MachidaHosokawa2013}, who model the formation of a protostar and follow its evolution for \sm$10^5$~yr until the envelope has dissipated\footnote{\citet{MachidaHosokawa2013} used a sink cell of 2~au and minimum cell size of 1.46~au, compared to our 1~au sink and minimum smoothing length in the disc of \sm0.2~au.}.  By the end of the Class 0 phase, they find discs with masses of 0.12-0.36~\Msun{}, which is a similar range to what we present in our study (at least for the models that form discs).  This suggests that discs gain most of their mass early on, although we would need to evolve our simulations considerably longer to conclusively prove this.  Thus, although we cannot comment on the long-term evolution based upon our models, we can and do describe the disc that forms simultaneously with the protostar, which appear to be a good representation of the Class 0 discs based upon the results of \citet{MachidaHosokawa2013}.

%----------------------------------------
\subsection{Comparison to other studies}
\label{sec:discs:comp}
There have been many studies that investigate the role of turbulence on star formation, and these studies collectively span a large parameter space.  Most studies involve cores of greater mass than we study here, but (as far as we know), this is the first study to include the three non-ideal  processes\footnote{\citet{WursterBatePrice2019} included both turbulence and non-ideal MHD, but their goal was to investigate the formation of a low-mass star cluster rather than an isolated star.}.  These studies have reached various conclusions, suggesting that the initial conditions are as important as the inclusion/exclusion of turbulence \citep[for a review, see][]{WursterLi2018}.  We note that when turbulence is included, the results (and possibly the conclusions) may differ simply as a result of the initial seed of the turbulent velocity field \citepeg{GoodwinWhitworthWardthompson2004a,Liptai+2017,Geen+2018}.  

By design, this study most closely resembles the ideal MHD study of \citet{LewisBate2018}, thus the results of our ideal simulations with \Bnz{} are in agreement with their conclusions that increasing turbulence hinders first core outflows\footnote{Although \citet{LewisBate2018} used \Bpz{}, ideal MHD is invariant to \Bpz{} $\rightarrow$ \Bnz{}, so our ideal results are directly comparable to theirs.}.  They also conclude that increasing turbulence (in  \Bpz{} models) hinders the formation of the pseudo-disc (the over-dense, disc-like structure with a radius of a few hundred au that appears in the mid-plane of laminar models; see their fig.~7), but do not comment on the formation of protostellar discs similar to those that we analyse here.  In agreement, we also find that increasing turbulence decreases the formation of these pseudo-discs (not shown), and this is true for all magnetised models with \Bpmz{}.
  
Our general conclusions are also in agreement with \citet{MatsumotoMachidaInutsuka2017} who used slightly more massive cores (2.5~\Msun) but also studied sub- and transsonic turbulence.  In their strongest magnetic field case, they found the level of turbulence did not affect the disc properties.   Therefore, in this parameter space (low-mass cores with sub- and transsonic turbulence), magnetic fields are the dominant process in influencing star and disc formation.

\citet{Hennebelle+2020} recently investigated the effect that magnetic field strength, misalignment between the magnetic field vector and rotation axis, initial rotation rate, and including transsonic turbulence had on the formation and evolution of discs.  Their models included ambipolar diffusion.  They concluded that, if the core is initially sufficiently magnetised (i.e. $\mu_0 \lesssim 6.7$), then the resulting disc masses and radii were relatively insensitive to their initial parameters.  This is in agreement with our work.  Given fixed initial physical properties, their disc masses and radii were sensitive to their sink size and the accretion scheme onto the sink, demonstrating the importance of carefully choosing numerical algorithms and parameters.  

\citet{GerrardFederrathKuruwita2019} initialised their 1~\Msun{} cores with a turbulent magnetic field rather than a turbulent velocity field.  By increasing the turbulent component of the magnetic field and decreasing the laminar \Bpz{} component, they found that first core outflows were inhibited.  Thus, first core outflows are inhibited when the uniform poloidal component of the magnetic field is disrupted, independent of whether this disruption is from an initially turbulent velocity, turbulent magnetic field or the \Bnx{} orientation.  

As the initial mass of the core is increased, so typically is the initial level of turbulence.  In these studies, the conclusions tend to have greater variance.  No discs formed in the 5~\Msun{} simulations of \citet{Joos+2013} and \citet{GrayMckeeKlein2018} that included a strong magnetic field (\mueq{2}); their initial Mach numbers spanned a range from sub- to supersonic.  When \citet{Joos+2013} decreased the initial magnetic field strength to \mueq{5} as in our study, low-mass discs formed in the subsonic models and high-mass discs formed in the supersonic models; when they decreased their initial magnetic field strength again, massive discs formed for all initial Mach numbers.  Therefore, these studies find that both magnetic field strength and turbulence are responsible for disc formation, where turbulence plays a dominant role at specific magnetic field strengths.

Larger mass cores \citep[$\geq50$~\Msun; ][]{Seifried+2013,GrayMckeeKlein2018,WursterBatePrice2019} are typically initialised with highly supersonic turbulence ($\mathcal{M}_0 \gtrsim 5$) to better match observations \citep{Larson1981}.  These clouds fragment into multiple stars, many of which have discs.  From their studies, \citet{Seifried+2013} and \citet{WursterBatePrice2019} concluded that in larger cores, (initially supersonic) turbulence was more important than magnetic fields in determining whether or not discs will form. \citet{WursterBatePrice2019} further concluded that this was independent of inclusion/exclusion of non-ideal MHD.  \citet{GrayMckeeKlein2018}, however, performed simulations where the total angular momentum vector resulting from the initial turbulent velocity field was initially aligned with the magnetic field, and found that no discs formed; when the total angular momentum vector was misaligned with the magnetic field, then they obtained one disc in each model (compared to three to 10 stars).  This suggests that the orientation of the total angular momentum vector resulting from turbulence is important, and that there must be a misalignment between the angular momentum vector and the magnetic field to form discs.  This conclusion regarding misalignment is reinforced by many studies who have investigated misaligned magnetic fields in laminar flows \citepeg{MatsumotoTomisaka2004,HennebelleCiardi2009,JoosHennebelleCiardi2012,KrumholzCrutcherHull2013,LewisBatePrice2015,Masson+2016,LewisBate2017}.

%----------------------------------------------------------------------------------------------------------------
\section{conclusion}
\label{sec:conclusion}

We numerically investigated the competing processes of sub/transsonic turbulence and magnetohydrodynamics on the formation of circumstellar discs  by modelling the gravitational collapse of 1~\Msun{} gas cores that are initially rotating and superimposed with a turbulent velocity field; in \citetalias{WursterLewis2020sc}, we investigate the effect these parameters have on the formation of the stellar core.  We tested two initial rotation rates, four initial Mach numbers and the inclusion/exclusion of ideal and non-ideal MHD with two to three directions of the initial magnetic field.

To study the formation and early evolution of circumstellar discs, we inserted 1~au sink particles late in the first hydrostatic core phase to represent the protostar and evolved the models for at least 1500~yr.  Our key results are as follows:
\begin{enumerate}
\item Discs preferentially formed in the purely hydrodynamic models and the non-ideal MHD models with \Bnx{} and \Bnz{}; when discs formed in the ideal MHD models, they were small and/or low mass.  This is independent of the initial level of turbulence.  Therefore, when employing ideal MHD, sub- and transsonic turbulence is not strong enough to prevent the magnetic braking catastrophe.
\item Increasing the initial Mach number had little impact on the disc properties (i.e. disc radius, mass, magnetic field strength) for the models that formed well-defined discs (i.e. the non-ideal MHD models with \Bnx{} and \Bnz{}).
\item For subsonic turbulence, rotation is more important than turbulence in purely hydrodynamical simulations; slowly rotating systems formed large discs, while quickly rotating subsonic systems fragmented to form multiple stars, many of which had discs.
\item First hydrostatic core outflows formed in the models with \Bpmz{} and were influenced by all of the initial conditions.  Increasing the Mach number inhibited the outflows, and suppressed them completely for transsonic turbulence.
\end{enumerate}

The initial level of turbulence influences the first core outflows.  For the remainder of the properties associated with circumstellar discs, non-ideal magnetohydrodynamical processes are more important than sub- and transsonic turbulent processes in determining the evolution of the system.

%----------------------------------------------------------------------------------------------------------------
\section*{Acknowledgements}

We would like to thank the anonymous referee for useful comments that greatly improved the quality of this manuscript.
JW acknowledges support from the European Research Council under the European Community's Seventh Framework Programme (FP7/2007- 2013 grant agreement no. 339248), and from the University of St Andrews.
BTL acknowledges the support of the National Aeronautics and Space Administration (NASA) through grant NN17AK90G and from the National Science Foundation (NSF) through grants no 1517488 and PHY-1748958.
The authors would like to acknowledge the use of the University of Exeter High-Performance Computing (HPC) facility in carrying out this work.  
Analysis for this work was performed using the DiRAC Data Intensive service at Leicester, operated by the University of Leicester IT Services, which forms part of the STFC DiRAC HPC Facility (www.dirac.ac.uk). The equipment was funded by BEIS capital funding via STFC capital grants ST/K000373/1 and ST/R002363/1 and STFC DiRAC Operations grant ST/R001014/1. DiRAC is part of the National e-Infrastructure.  
The column density figures were made using \textsc{splash} \citep{Price2007}.
%The research data supporting this publication are openly available from the University of Exeter's institutional repository at: https://doi.org/10.24378/exe.XXX.
The research data supporting this publication will be openly available from the University of Exeter's institutional repository.

%----------------------------------------------------------------------------------------------------------------
\bibliography{TurbVsNimhd_discs}
%----------------------------------------------------------------------------------------------------------------

\label{lastpage}
\end{document}